\documentclass[reprint,showpacs,amsmath,amssymb,superscriptaddress, longbibliography]{revtex4-1}
\usepackage{graphicx}
\usepackage{dcolumn}
\usepackage[usenames,dvipsnames]{xcolor} 
\usepackage{changes}
\usepackage{hyperref}

\makeatletter
\newcommand*{\rom}[1]{\expandafter\@slowromancap\romannumeral #1@}

\begin{document}

\title{Microelectromechanical system based design of a high-finesse fiber cavity integrated with an ion trap}

\author{Moonjoo Lee}
\thanks{These authors contributed equally to this work.}
\affiliation{Institut f\"ur Experimentalphysik, University of Innsbruck, Technikerstra\ss e 25, 6020 Innsbruck, Austria}
\author{Minjae Lee}
\thanks{These authors contributed equally to this work.}
\affiliation{ISRC/ASRI, Department of Electrical and Computer Engineering, Seoul National University, Seoul, 151-744, Korea}
\author{Seokjun Hong}
\affiliation{ISRC/ASRI, Department of Electrical and Computer Engineering, Seoul National University, Seoul, 151-744, Korea}
\author{Klemens Sch\"uppert}
\affiliation{Institut f\"ur Experimentalphysik, University of Innsbruck, Technikerstra\ss e 25, 6020 Innsbruck, Austria}
\author{Yeong-Dae Kwon}
\affiliation{Quantum Technology Laboratory, SK Telecom, Seongnam-si, 463-784, Korea}
\author{Taehyun Kim}
\affiliation{ICT/ASRI, Department of Computer Science and Engineering, Seoul National University, Seoul, 151-744, Korea}
\author{Yves Colombe}
\affiliation{Institut f\"ur Experimentalphysik, University of Innsbruck, Technikerstra\ss e 25, 6020 Innsbruck, Austria}
\author{Tracy E. Northup}
\thanks{tracy.northup@uibk.ac.at}
\affiliation{Institut f\"ur Experimentalphysik, University of Innsbruck, Technikerstra\ss e 25, 6020 Innsbruck, Austria}
\author{Dong-Il ``Dan'' Cho}
\thanks{dicho@snu.ac.kr}
\affiliation{ISRC/ASRI, Department of Electrical and Computer Engineering, Seoul National University, Seoul, 151-744, Korea}
\author{Rainer Blatt}
\affiliation{Institut f\"ur Experimentalphysik, University of Innsbruck, Technikerstra\ss e 25, 6020 Innsbruck, Austria}
\affiliation{Institut f\"ur Quantenoptik und Quanteninformation, \"Osterreichische Akademie der Wissenschaften, Technikerstra\ss e 21a, 6020 Innsbruck, Austria}


\begin{abstract}
We present a numerical study of a MEMS-based design of a fiber cavity integrated with an ion trap system. 
Each fiber mirror is supported by a microactuator that controls the mirror's position in three dimensions.
The mechanical stability is investigated by a feasibility analysis showing that the actuator offers a stable support of the fiber.
The actuators move the fibers' positions continuously with a stroke of more than 10~$\mu$m, with mechanical resonance frequencies on the order of kHz. 
A calculation of the trapping potential shows that a separation between ion and fiber consistent with strong ion-cavity coupling is feasible.
Our miniaturized ion-photon interface constitutes a viable approach to integrated hardware for quantum information.
\end{abstract}

\maketitle

\section{Introduction}


The trapped ion system is one of the most promising candidates for quantum information processing~\cite{Monroe-Science13, Schindler-NJP13}.
Recent progress includes the generation of entanglement across 20 individually controlled ions~\cite{Friis-PRX18}, memory times exceeding 10 minutes~\cite{Wang-NatPhot17}, and a single-qubit gate fidelity of 99.9999\%~\cite{Harty-PRL14}. 
When these advantages are combined with those of optical cavities, the ion-cavity setting can be used to realize classical simulations~\cite{Bylinskii-Science15, Gangloff-NatPhys15, Bylinskii-NatMat16} and is expected to enable efficient quantum networks~\cite{Moehring-Nature07, Northup-NatPhot14}.
Moreover, the cavity field mediates  a long-range interaction between ions, which makes the system a suitable platform to study many-body physics such as structural phase transitions~\cite{Fogarty-PRL15, Rojan-PRA16}.


In order to demonstrate quantum advantage with ion-based quantum devices, that is, to solve classically intractable problems~\cite{Preskill-arXiv12}, it will be necessary to scale up traps to store and manipulate tens or hundreds of qubits.
Coupling such ion traps to optical cavities, we can develop advanced quantum information systems, such as a quantum repeater node comprising memory and communication qubit ions~\cite{Monroe-Science13} and a modular quantum computer distributed across a photonic network~\cite{Monroe-PRA14, Kim-PRA11}.
To address the challenge of scalability, one solution is to employ microelectromechanical systems (MEMS) fabrication technology to miniaturize an ion-cavity system.


MEMS-based ion trap technology is regarded as a promising platform to build large-scale quantum systems~\cite{Kielpinski-Nature02, Cho-MNSL15}. 
The first MEMS surface ion trap was fabricated at NIST Boulder in 2005~\cite{Chiaverini-QIC05} by evaporating Au electrodes on fused silica. 
A MEMS surface ion trap has smaller trap depth than a three-dimensional Paul trap~\cite{Brownnutt-NJP06} but has the advantage of a reconfigurable planar trapping geometry, along with extensive optical access for laser beams, as required for a large-scale ion trap quantum computer. 
In 2016, a MEMS trap called the High Optical Access 2.0 trap was developed by Sandia National Laboratory~\cite{Maunz-HOA16}, which is widely used by many ion trap researchers today. 
Topics of active research include the question of how to reduce stray charge accumulation, e.g., on the trap sidewalls~\cite{Hong-JMS18} or via \textit{in situ} cleaning~\cite{Hite2012, Hite2013}, and how to build increasingly sophisticated structures, e.g., junction traps~\cite{Amini-NJP10, Moehring-NJP11} for ion transport, and two- or three-dimensional electrode arrays~\cite{Seidelin-PRL06, Wilpers-NatNano12}.


MEMS-based ion traps have advantages in scalability and ease of fabrication, and they can also be easily integrated with other type of MEMS devices.
MEMS techniques can also reduce the footprint of the optical cavity system~\cite{Gollasch-JMM05, Srinivasan-SAA10}, particularly for the mirror actuators.
The replacement of standard high-finesse mirrors by fiber mirrors has already reduced the physical cavity volume significantly~\cite{Hunger-NJP10, Takahashi-PRA17}; however, when commercial nanopositioning stages are used, they place significant space demands on the in-vacuum assembly. 
Here, we propose and investigate a novel design for a MEMS-based fiber-cavity system integrated with a surface ion trap. 
The fiber system is studied by analyzing the mechanical stability, resonances, and stroke.
Furthermore, we calculate the trapping potential seen by the ions in order to discuss the prospects for strong ion-cavity coupling.


\begin{figure*} [!t]
\includegraphics[width=7.0in]{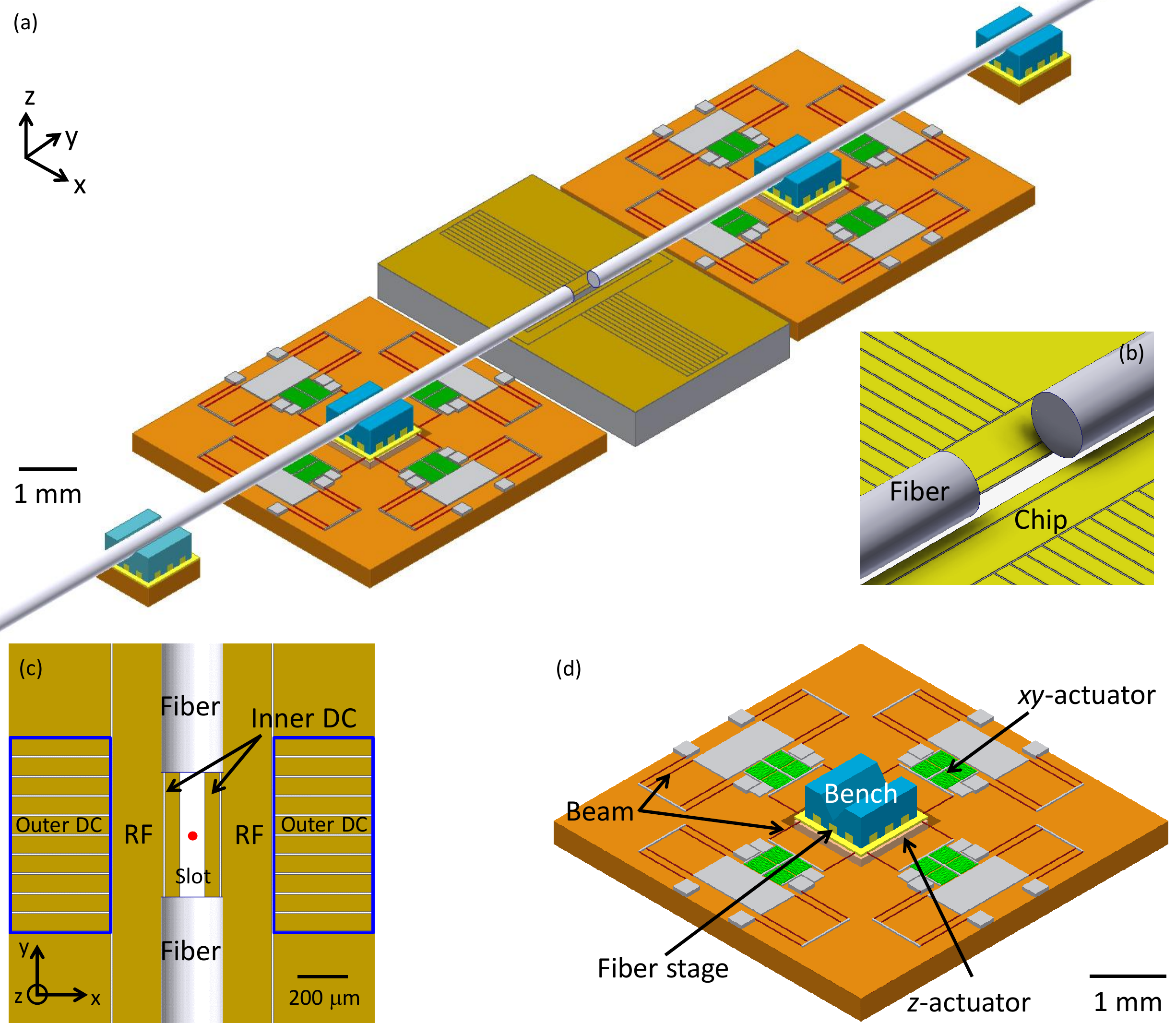} 
\caption{
(a) Overview of our design for a MEMS-based fiber cavity integrated with a surface ion trap.
The central piece is the trapping chip, and the structures on both sides correspond to the fiber cavity system.
(b) Zoomed-in view of the ion trapping site. 
(c) Electrode configuration of the chip, in which the RF rails, outer DC electrodes, and inner DC rails are indicated. 
Confinement along the cavity axis is achieved by voltages on the outer DC electrodes, marked by blue boxes. 
The red dot in the center indicates the ions' position.
The gap between neighboring electrodes is 8~$\mu$m. 
(d) Detailed view of the fiber stage and actuators.
A set of actuators control the position of the stage in three dimensions via beams.
A fiber will be glued onto the V-shaped fiber bench.
}
\label{fig:overview}
\end{figure*}


\begin{figure*} [!t]
\includegraphics[width=6.0in]{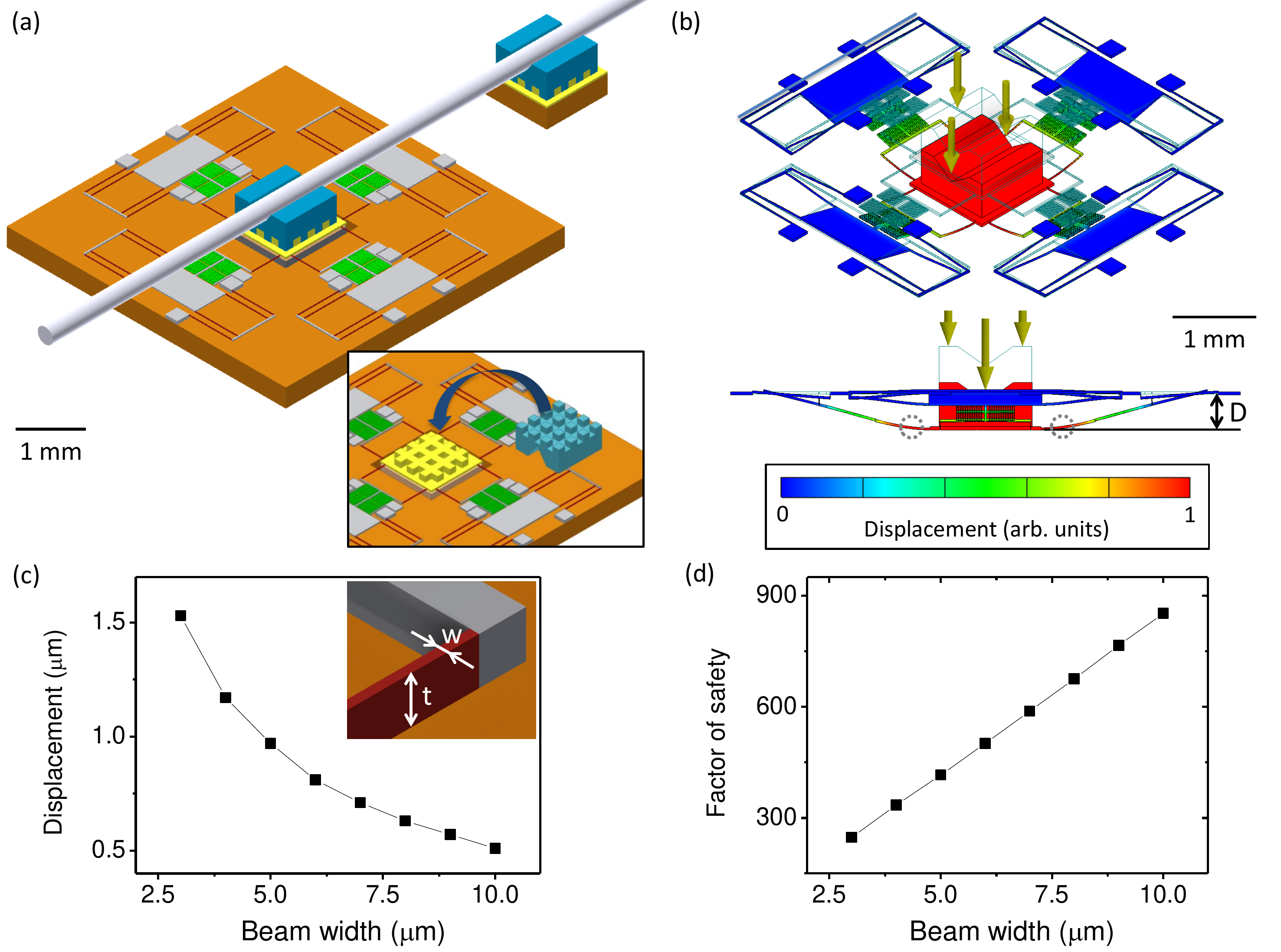} 
\caption{
Fiber stage and stability analysis.
(a) Schematic of the fiber stage with the fiber installed on the fiber bench. 
A second supporting stage can be used to relieve the load of the primary fiber stage. 
Inset: Plug-and-socket structures define the position of the bench with respect to the stage.
(b) Equilibrium position of the stage, which is slightly shifted down due to gravity (fiber not shown).
The gray dotted circles indicate the position of the minimum FOS in the whole structure.
(c) Vertical displacement (``D" in (b)) as a function of beam width \textit{w}, indicated in the inset.  The beam thickness \textit{t}, also indicated, is fixed at 20~$\mu$m for all simulations in this paper.
(d) Factor of safety as a function of beam width. 
}
\label{fig:stress}
\end{figure*}

\section{Basic concept}


The main structure consists of a surface ion trap integrated with a MEMS-based fiber cavity system (Fig.~\ref{fig:overview}(a)).
The starting point is a microfabricated chip, detailed views of which are found in  Figs.~\ref{fig:overview}(b) and (c), and which has previously been described in Ref.~\cite{Cho-MNSL15}. 
The chip has a linear Paul trap configuration in which two radio-frequency (RF) rails are used to generate a pseudopotential tube along the $y$ axis. 
Axial confinement of ions is achieved by applying voltages to outer DC electrodes. 
The axis of the ion string is designed to align with that of the cavity such that multiple ions can be coupled to the cavity mode. 
Two inner DC rails define the principal axes of the trapping potential to ensure that the $k$-vector of the Doppler cooling laser has overlap with all axes.  
The ion height above the trap surface may be controlled by adding RF voltages to the inner DC electrodes~\cite{Boldin-PRL18}.
Between the two inner DC electrodes, a slot allows for optical access to the ions from beneath the chip.


A cavity is formed by two fiber mirrors~\cite{Hunger-NJP10}, each supported by a MEMS-based fiber stage, for which actuators offer three-dimensional position control, as shown in Figs.~\ref{fig:overview}(a) and (d).
A closer view of the stage is shown in Fig.~\ref{fig:stress}(a).
A V-shaped bench holding a fiber is fixed to each fiber stage, while the stage itself is suspended by thin beams, the other ends of which are connected to the surrounding actuators.  
Position control of the fiber stage is based on the mechanism of electrostatic actuation~\cite{Tang-SA89, Legtenberg-JMM96, Gollasch-JMM05}: As described in detail in Sec.~\rom{3}, a comb drive actuator provides continuous motion in the $xy$ plane.
The fiber stage can be moved along the $z$ axis via the electrostatic force between the stage and the fixed plate below.
Such continuous positioning is intended to enable us to compensate for misalignment between the fiber mirrors, to align the cavity mode to the ion position, and to scan or lock the cavity frequency. 
The trapping chip and the fiber stages are fixed together using a common base plate with position-defining grooves. 
The fiber stages and trapping chip are glued individually to the base plate.


\begin{figure*} [!t]
\includegraphics[width=6.5in]{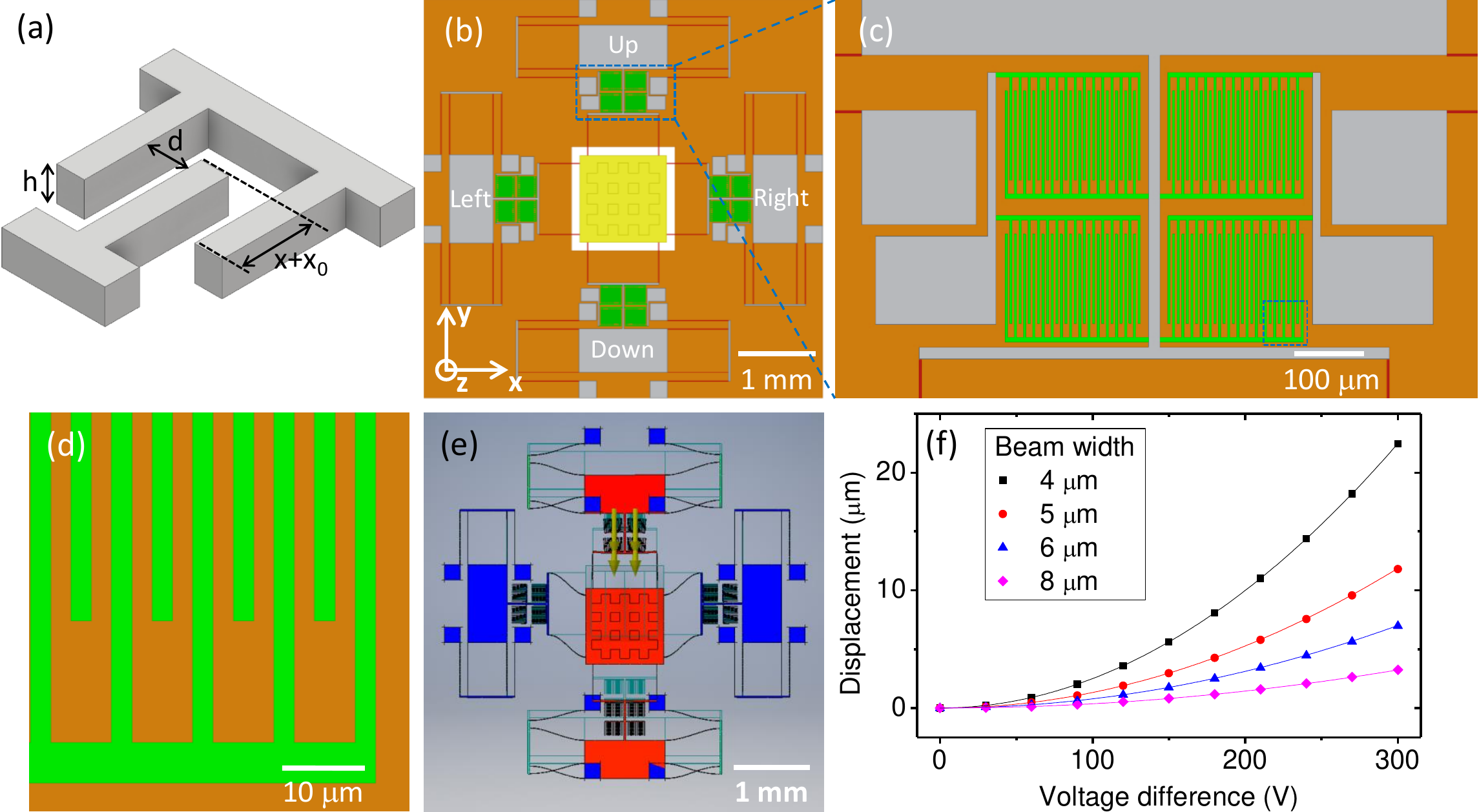} 
\caption{
(a) A single cell of the comb drive, with the thickness of the fingers $h$, the length of the gap between the two parts perpendicular to the displacement axis $d$, the displacement $x$, and the overlappling length between the two parts along the displacement to the displacement axis without a voltage difference $x_0$.
A voltage difference between the two parts changes their relative distance via the electrostatic force.
(b) Top view of the fiber stage (center) with the plug-and-socket structure indicated. 
The stage is suspended from the comb drive via two beams on each side.
Both ``Left'' and ``Right'' comb drives  move the stage along the $x$ axis, and the comb drives denoted as ``Up'' and ``Down'' are used for the positioning along the $y$ axis.
(c) Closer view of the ``Up'' comb structure.
(d) Closer view of the blue box in (c).
(e) Exaggerated displacement along the $y$ axis.
The beams in the direction orthogonal to the movement are deflected.
(f) The stage displacement along the $y$ axis is calculated as a function of the driving voltage for beam widths of 4~$\mu$m (black squares), 5~$\mu$m (red circles), 6~$\mu$m (blue triangles), and 8~$\mu$m (magenta diamonds). 
The solid lines are obtained by fitting with a quadratic function. 
}
\label{fig:comb}
\end{figure*}

\section{Fiber stage and actuator}

\subsection{Mechanical stability}

\label{fiber-stage-and-actuator}
The mechanical stability of the fiber stage is investigated to check whether the stage can provide a safe support~\cite{Tang-JMS92}.
All numerical simulations for investigating mechanical properties are performed with COMSOL Multiphysics. 
We consider a fiber of SiO$_2$ with diameter 250~$\mu$m glued on the fiber bench.
The bench, the fiber stage, and the actuators are made of Si~\cite{Diem-SAA95, Renard2000, Kim-Trans03, Lee-JMS99}.
The load for the stage is conservatively estimated as 1.8 mg, including the fiber~\cite{MEMS-SI18}.
The mass and volume of the glue are neglected~\cite{Goettsche-IJAMT07}.
First, we numerically model the influence of the weight on the vertical stage position, with the results shown in Figs.~\ref{fig:stress}(b) and (c). 
The load slightly shifts down the stage level due to gravity: The level of the stage descends by 0.97~$\mu$m for a beam width of 5~$\mu$m.
The beams correspond to the dark brown bridges indicated in Fig.~\ref{fig:overview}(d). 
Each $xy$-actuator possesses 40 beams so that one fiber stage is controlled through 160 beams.
For the calculations in Figs.~\ref{fig:stress}(c) and (d), the width of all 160 beams is changed.
Second, the factor of safety (FOS) is calculated.
The FOS is defined as the yield stress divided by the working stress and has been commonly used to evaluate the mechanical stability of structures~\cite{Miller-FOS10}.
A system is considered to be a stable structure if the FOS is larger than 3. 
In our system, the FOS is minimized at the position of the largest curvature within the beams directly connected to the fiber stage (gray dotted circles in Fig.~\ref{fig:stress}(b)).
In Fig.~\ref{fig:stress}(d), we plot the minimum FOS as a function of the beam width.
The obtained FOS values for several beam widths indicate that the stage is stable enough to support the fiber.
Note that a fiber feedthrough will be used to bring the fiber into the vacuum chamber in which the system will be mounted.  
Tension between the fiber and the fiber feedthrough will exert an additional mechanical stress.
This stress can be relieved by careful positioning of the fiber stage or by inserting a strain-relieving stage, for instance, a piezoelectric element, between the fiber and the feedthrough.


\begin{figure} [!t]
\includegraphics[width=3in]{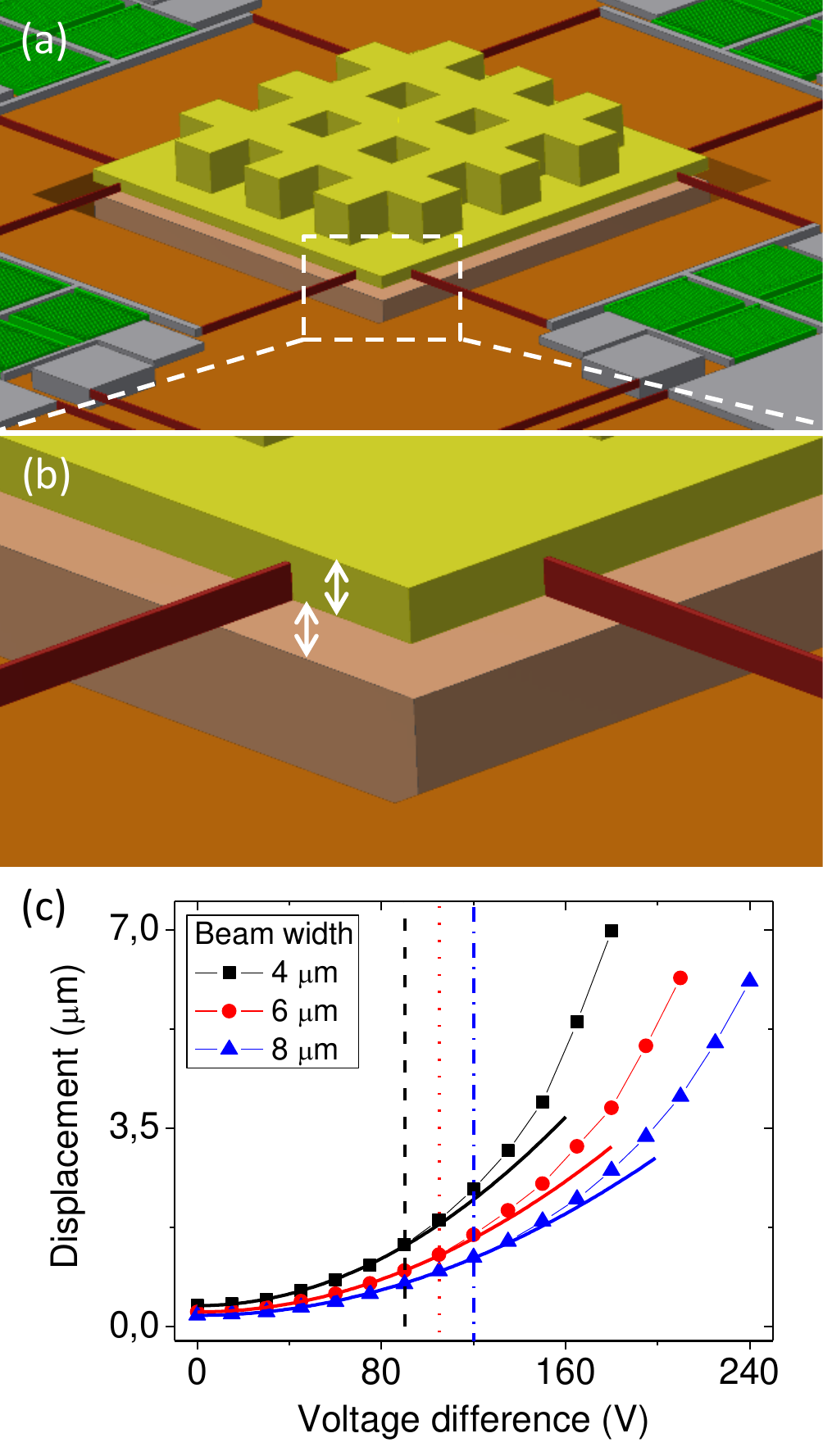} 
\caption{
The actuator for movement along the $z$ axis. 
(a), (b) Views of the fiber stage (yellow) and the fixed plate underneath (grey).
The thickness of the stage and the gap between the stage and the plate are both 30~$\mu$m (white arrows in (b)). 
A voltage difference between the stage and plate displaces the stage along the $z$ axis via the electrostatic force.
(c) Stage displacement along the $z$ axis as a function of the voltage difference for a beam width of 4~$\mu$m (black squares), 6~$\mu$m (red circles), and  8~$\mu$m (blue triangles).
The solid lines are fit to quadratic functions for comparison.
The fitting range is from 0 V to the vertical black dashed line for 4~$\mu$m, red dotted line for 6~$\mu$m, and blue dotted dashed line for 8~$\mu$m, for which V$^2$ scaling holds.
}
\label{fig:zscan}
\end{figure}

\subsection{Positioning in three dimensions}


We control the position of the stage in the $x$ and $y$ directions with comb-drive actuators using the electrostatic force~\cite{Legtenberg-JMM96, Liu-JMM07, Maroufi2016}.
Two- and three-dimensional comb-drive actuators have been demonstrated in various systems such as microgrippers, atomic force microscopes, and scanning tunneling microscopes~\cite{Indermuhle-SAA95, Kim-JMS92, Ando-SAA04, Maroufi-JMS17}.
A single unit of the comb drive is shown in Fig.~\ref{fig:comb}(a) to describe the working principle.
The comb drive has two parts: a moving part and a stationary part. 
The capacitance between the two parts is given by 
\begin{equation}
C(x) = \frac{2N\epsilon_{0}h}{d}(x+x_{0}),
\end{equation}
where $x$ is the displacement, $x_0$ is the overlapping length between the two parts along the displacement axis without a voltage difference, $N$ is the total number of the comb units, $h$ is the thickness of the fingers, $\epsilon_{0}$ is the dielectric constant of the vacuum, and $d$ is the length of the gap between the two parts perpendicular to the displacement axis. 
If a voltage difference $V$ is applied between the two parts, the exerted force can be expressed as
\begin{equation}
F = \frac{1}{2}\frac{\partial C}{\partial x}V^{2}= \frac{N\epsilon_{0}h}{d}V^{2},
\label{eg:comb_force}
\end{equation}
i.e., the force is proportional to the change of the capacitance between two comb fingers.
Here we assume that $x_{0}$ is large enough that side effects like a fringe field can be neglected. 
This force results in a displacement of the moving part of
\begin{equation}
x = \frac{N\epsilon_{0}h}{kd}V^{2}, 
\label{eg:comb_displacement}
\end{equation}
with the spring constant $k$ of the structure.


Our comb-drive configuration is shown in Figs.~\ref{fig:comb}(b)-(e).
The comb drives indicated as Up and Down in Fig.~\ref{fig:comb}(b) induce motion along the $x$ axis, and the drives indicated as Left and Right induce $y$-axis motion. 
Each comb drive has 120 teeth, such that in total the force of 240 teeth drives the motion along a given axis. 
As visualized in Fig.~\ref{fig:comb}(b), if the position of a moving part is shifted, the force is transmitted between the comb drive and the fiber stage via beams on each side.
Note that the beams in the transverse directions are deflected by this motion. 
Fig. 3(f) shows the simulation result of the stroke as a function of driving voltage for three different values of beam widths. 
As the width decreases, the beam becomes more elastic, which corresponds to a decreasing spring constant $k$.
Hence, for a given voltage, the displacement is larger if the beam is thinner.
As expected from Eq.~\ref{eg:comb_displacement}, this displacement follows a quadratic scaling with respect to the driving voltage.
One main benefit of MEMS actuators as compared to piezoelectric actuators consists in the larger stroke. 
In our case, for a beam width of 4~$\mu$m, we can cover one free spectral range of the fiber cavity (about 400 nm) with only 20 V.
We also check the mechanical stability for the displacement along the $x$ and $y$ axes. 
As the displacement increases, the working stress to the actuator also increases, resulting in the reduction of the FOS. 
Our calculated FOS values at a maximum applied voltage of 300 V correspond to 13, 21, 32, and 61 for beam widths of 4, 5, 6, and 8 $\mu$m, respectively, showing that our actuator is stable under the maximum displacement.

\begin{figure} [!b]
\includegraphics[width=3.5in]{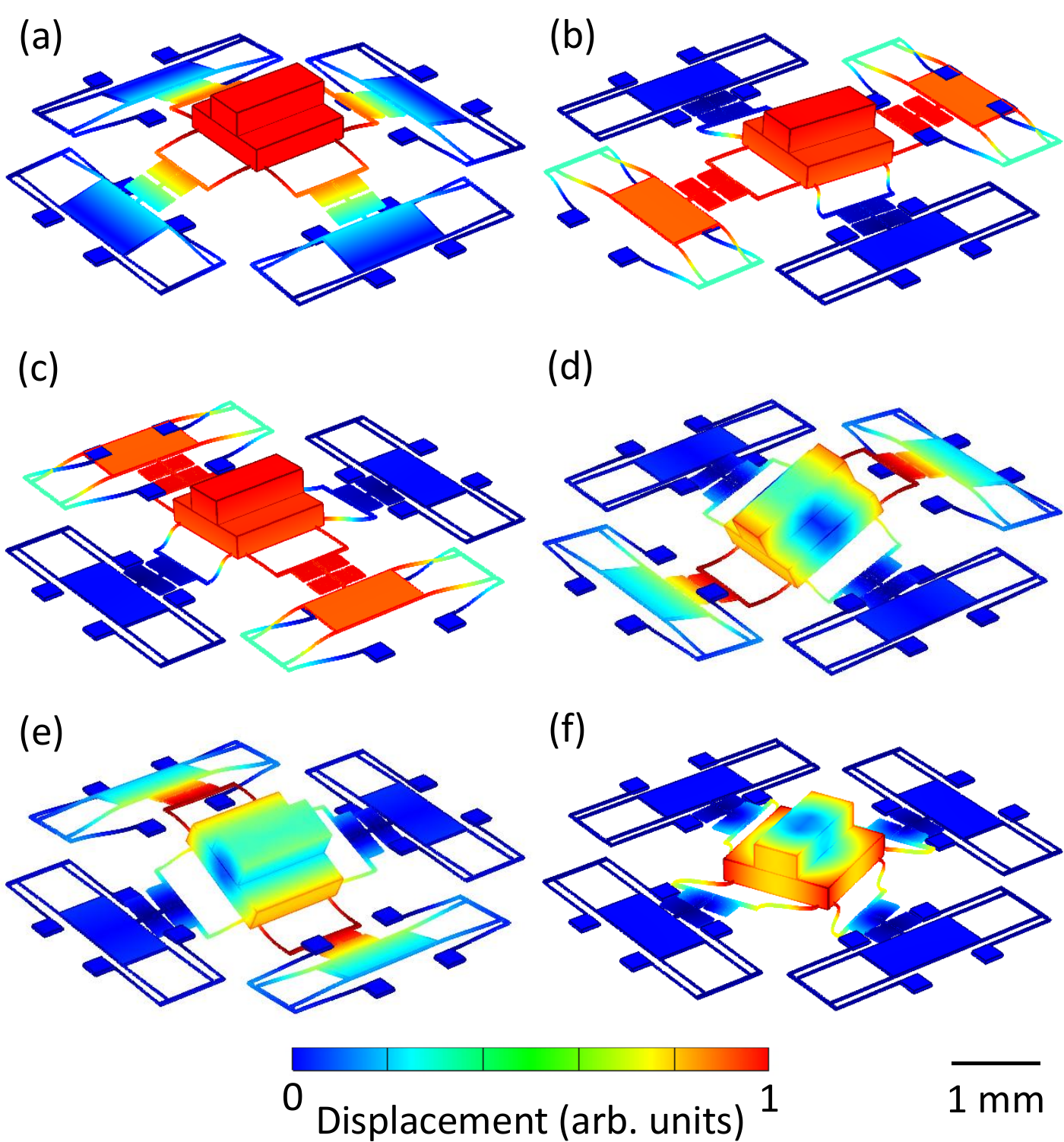}
\caption{
Simulated mechanical modes of the fiber stage for a beam width of 8~$\mu$m, ordered by ascending frequency.
The resonance frequencies without and with fibers are calculated as:
(a) 985 Hz and 633 Hz; (b) 1.53 kHz and 952 Hz; (c) 1.62 kHz and 963 Hz; (d) 3.37 kHz and 2.15 kHz; (e) 3.58 kHz and 2.82 kHz; (f) 7.27 kHz and 4.87 kHz. 
}
\label{fig:resonance}
\end{figure}


\begin{figure*} [!t]
\includegraphics[width=6.5in]{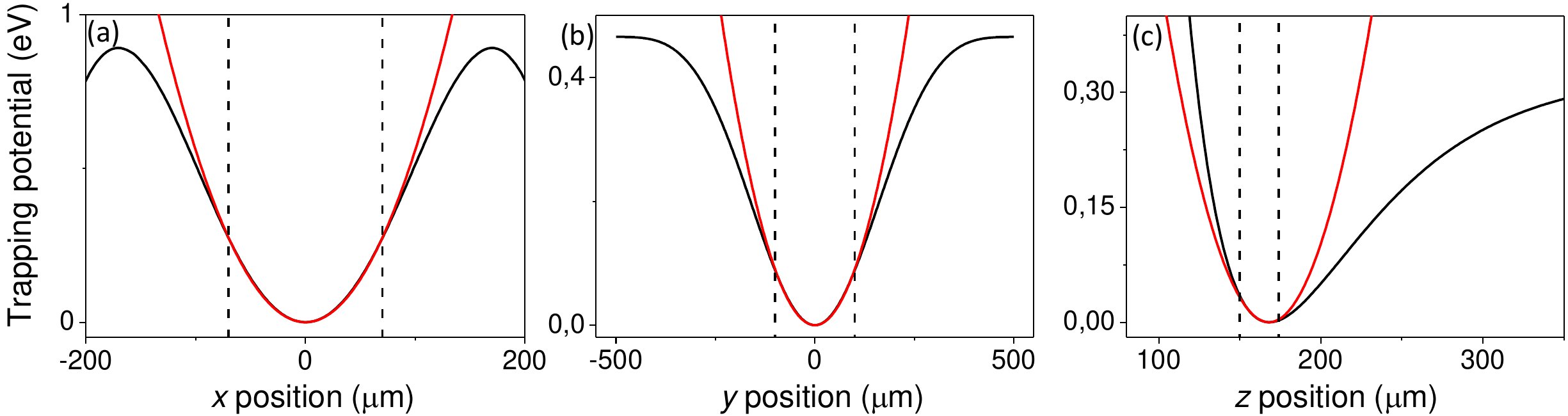} 
\caption{Simulated trapping potential profiles for a cavity of length 1 mm comprised of uncharged and uncoated fiber mirrors. 
The black lines are the results of calculations, and the red lines are obtained by fitting the data with quadratic functions. 
The fitting is done for the data between two vertical lines.
The profile is shown along $(x, 0, 0)$ in (a), $(0, y, 0)$ in (b), and $(0, 0, z)$ in (c).
The origin is the center of the trapping chip surface.
}
\label{fig:profile}
\end{figure*}


The position of the fiber stage along the $z$ axis is controlled with the electrostatic force between the stage and another plate fixed below (see Figs.~\ref{fig:zscan}(a) and (b)).
This actuator for the $z$ axis also makes use of a position-dependent capacitance.  
In Fig.~\ref{fig:zscan}(c), the simulation result shows the vertical displacement with respect to the driving voltage for three beam widths.
It is notable that here, too, we obtain large strokes: For a beam width of 4~$\mu$m, the stage is displaced by 5~$\mu$m with an applied voltage of 160 V. 
The main difference from the comb drive is that when a voltage is applied, both the force and the distance between two plates change simultaneously: The distance $d$ in Eq.~\ref{eg:comb_force} is also a function of $V$, resulting in a deviation from the $V^2$ scaling (Fig.~\ref{fig:zscan}(c)).
To check the mechanical stability, the FOS is calculated at the maximum applied voltage for each beam width. 
For beam widths of 4, 6, and 8 $\mu$m, we obtain FOS values of 81, 95, and 99 while voltage differences of 180, 210, and 240 V are applied for each width.
These calculation results demonstrate the safety of the $z$-axis motion.
More complicated movements could be controlled by segmenting the electrodes on the lower plate. 
Separately applied voltages on these electrode segments would apply torque to the fiber stage, enabling rolling or pitching motion.

\subsection{Mechanical resonances}


The mechanical modes of the fiber stage and their resonance frequencies are calculated to estimate the mechanical bandwidth.
For stabilizing the cavity frequency, it is desirable to achieve high mechanical bandwidth since it determines the maximum noise frequencies that can be suppressed~\cite{Briles-OE10}.
This bandwidth is typically limited by the mechanical resonances of the actuator: Phase shifts close to the resonance, accompanied by an amplitude response, would cause instabilities in the frequency stabilization~\cite{Gallego-APB16}.
Our simulation results are shown in Fig.~\ref{fig:resonance} for a beam width of 8~$\mu$m.
We find the lowest mechanical frequency of the stage at 985 Hz without the fiber (Fig.~\ref{fig:resonance}(a)) and investigated the six lowest mechanical resonances up to 7.27 kHz. 
With the fiber on top, the resonance frequency decreases due to the mass of the fiber with the lowest frequency at 633 Hz and the sixth at 4.87 kHz.

The mechanical resonance frequencies of our system are higher than those of commercially available stages 
but lower than those of recently reported fiber-cavity transducers~\cite{Brachmann-OE16, Gallego-APB16, Janitz-OE17}.
If the resonances of our actuator are problematic for locking the cavity frequency in a future experiment, three methods can be used to improve the performance of the cavity lock: The first method consists of applying a feedback signal to the piezo transducer of the optional secondary mount (Fig.~\ref{fig:stress}(a)), which would extend the bandwidth to the standard level of the piezoactuators of several kHz. 
Another approach is to utilize the photothermal effect, using which a locking bandwidth of 400 kHz has been demonstrated in a fiber cavity setting~\cite{Brachmann-OE16}. 
Finally, the phase lag in the vicinity of the resonance can be electronically compensated through digital filtering systems~\cite{Ryou-RSI18}.

\section{Trapping potentials}


Next, we investigate the trapping potential for the ions. 
The total potential $\phi(\bm{r})_{\rm{total}}$ is given by the sum of the pseudopotential generated by an RF voltage applied to the RF rails and the static potential $\phi(\bm{r})_{\rm{DC}}$ derived from voltages on the outer DC electrodes and from charges on the fibers:
\begin{equation}
\phi(\bm{r})_{\rm{total}}=\frac{q}{4m\Omega^2}|\vec{E}(\bm{r})|^2 + \phi(\bm{r})_{\rm{DC}},
\end{equation}
where $q$ and $m$ indicate the charge and mass of a single ion, respectively.
The driving RF frequency is chosen to be $\Omega =20$ MHz, and $\vec{E}(\bm{r})$ refers to the electric field due to the RF voltage.
The calculation is done for $^{171}$Yb$^+$ ions using COMSOL Multiphysics.
The RF amplitude is fixed at 150 V for all simulations of the trapping potentials, and DC voltage values for each simulation are given in~\cite{MEMS-SI18}.


We characterize the trapping potential using two parameters: trap depth and trap frequency.
The trap depth is defined as the potential difference between the minimum (at the trapping site) and the first local maximum of the profile along the $x$, $y$, and $z$ axes. 
The trap frequency along each axis is obtained by fitting the potential with a quadratic function. 
In order to exclude anharmonic regions of the potential profile, the fitting range is limited to $\pm70$ $\mu$m along the $x$ axis, to $\pm 100$ $\mu$m along the $y$ axis, and from -18 $\mu$m to +6 $\mu$m along the $z$ axis.
Here, the trapping potential is calculated for three cases: (1) uncharged and uncoated fiber, (2) uncharged fiber with metal coating on the sidewall, and (3) charged fiber with metal coating.

\subsection{Influence of fiber dielectrics}


We begin with how the dielectric surface and volume of the fiber mirrors affects the trapping potential, assuming that the fibers are uncoated and have a neutral charge.
In Fig.~\ref{fig:profile}, trapping potential profiles are presented along the $x$, $y$, and $z$ axes for a cavity length of 1 mm. 
The trap frequencies obtained from fits to quadratic functions are $(\omega_{x}, \omega_{y}, \omega_{z})/2\pi=(3.0, 1.1, 2.4)$ MHz~\cite{MEMS-SI18}.  
The deviation of each profile with respect to that of a harmonic potential is below~$<5\%$ for $|x|<86~\mu$m, $|y|<180~\mu$m, and $-22~\mu$m$<z<5~\mu$m.
The trap depth is $0.71$ eV, $0.46$ eV, and $0.33$ eV in each direction. 
Figs.~\ref{fig:potential}(a)-(b) show the ion height above the trap and the trap frequency as a function of the cavity length.
The trapping potentials are obtained in a step of 50 $\mu$m for a cavity length from 400 $\mu$m to 2000 $\mu$m.
The height of the fibers' centers is fixed at 169~$\mu$m, overlapping with the height of the ion above the trap in the absence of the fibers.
As the fibers are brought closer to the trapping site in our simulations, we find that the fibers have only a small influence on the potential for cavity lengths up to 500~$\mu$m.
For this cavity length, we obtain trap frequencies of $(3.0, 1.0, 2.5)$ MHz and find that the ion height is unchanged.
Therefore, if charges are absent from the fibers or can be removed by a discharging method~\cite{Pollack-PRD10}, this bare fiber becomes an option to make a short cavity offering strong ion-cavity coupling (see Sec.~\rom{5}).

\subsection{Influence of metal coating}
\label{dielectrics-metal}

\begin{figure} [!t]
\includegraphics[width=3.2in]{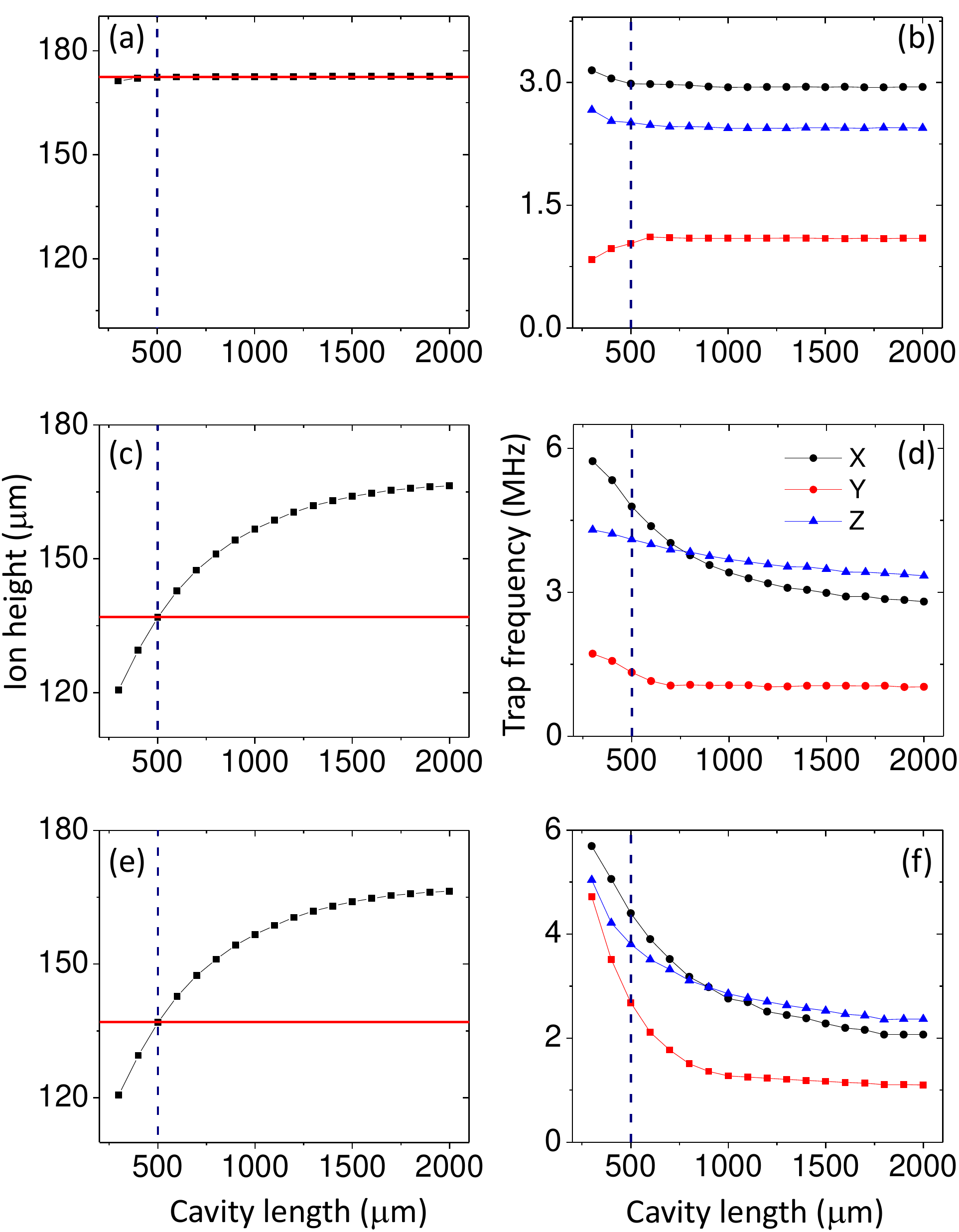} 
\caption{
Trapping potential simulations for cavity lengths from 300~$\mu$m to 2.0 mm. 
Ion height above the trap surface and trap frequency are plotted as a function of the cavity length for uncharged and uncoated fibers in (a) and (b), uncharged and metal-coated fibers in (c) and (d), and charged and metal-coated fibers in (e) and (f).
The vertical navy lines indicate a cavity length of 500~$\mu$m, at which the strong ion-cavity coupling condition can hold (see Sec.~\rom{5}). 
In (a), (c), and (e), the horizontal red line indicates the height of the fiber center, which has been chosen to overlap with the height of an ion for a cavity length of 500~$\mu$m.
}
\label{fig:potential}
\end{figure}

We consider a situation in which the fibers' sidewalls are coated with a thin metal film, as has been used to provide an ultra-high-vacuum-compatible coating in fiber-cavity experiments with neutral atoms and trapped ions~\cite{Colombe-Nature07, Hunger-NJP10, Steiner-PRL14, Podoliak-PRAppl16}.
This coating has two additional roles, along with protecting the fiber: The area of dielectric surface exposed to the ion is reduced, and a voltage can be applied to the fiber coating to compensate for distortion caused by charges on the dielectrics. 
In Figs.~\ref{fig:potential}(c)-(d), the simulation result is shown for a grounded metal coating and uncharged fiber facets. 
As the fibers are brought closer to the trap center, the height above the trap surface at which the potential minimum occurs goes down, and the trap frequency increases. 
This behavior can be interpreted by considering grounded metal boundaries approaching the trapping site from far away in all directions.
The potential energy becomes more confined as the metal structures come closer, leading to higher trap frequencies for shorter cavities. 
The effect is spatially symmetric along the $x$ and $y$ axes; however, the trapping chip makes it asymmetric along the $z$ axis.
In that case, one can consider a ground metal boundary approaching from above and shifting down the position of the ion height.


The fiber height is fixed at 138~$\mu$m for all cavity lengths so that it coincides with the height of a trapped ion for a 500~$\mu$m long cavity.
This height is found in the following way: Fixing the cavity length at 500~$\mu$m, the trapping potential is calculated for fiber heights from 169~$\mu$m to 130~$\mu$m with a step of 1~$\mu$m, so that we find out the fiber height that corresponds to the ion height.
Our calculation shows that the trapping potential for a 500~$\mu$m long cavity has frequencies of $(4.8, 1.3, 4.1)$ MHz, and the most shallow trap depth is 0.29 eV in the $+z$ direction. 
Note that the RF contribution to the potential along the $y$ axis is negligible with bare fibers.
In contrast, here the contribution of the RF psuedopotential is nonnegligible along the $y$ axis since the metal coating plays a role in determining the $y$-axis boundary conditions. 
This contribution would not affect micromotion compensation for a single ion but would make compensation difficult for a long ion string.


\subsection{Influence of charges}


Next, we investigate the influence of fiber charges on the trapping potential.
As in Sec.~\ref{dielectrics-metal}, we assume that the fiber sidewalls are metal coated. Here, we make the additional assumption that both fiber facets are positively charged with a surface density of 5~$q/\mu$m$^2$. 
This assumption is based on a measurement in our ion-fiber cavity setting at the University of Innsbruck: Approaching a fiber to single ions for ion-fiber distances from 1.6 mm to 200 $\mu$m, we measured the ions' positions and trap frequencies, affected by the charges on the fibers, as a function of the ion-fiber distance.
Reconstruction of the charge densities shows that the fibers are positively charged with densities of on the order of several elementary charges in an area of 1 $\mu$m$^2$~\cite{Ong-NJP18}. 
In the current simulation, we choose the charge density of 5~$q/\mu$m$^2$ that is very close to a median of the reconstructed densities.
The simulation result is shown in Figs.~\ref{fig:potential}(e)-(f).
The ion heights are similar to those found with an uncharged metal coating, and the fiber height is also fixed here at 138~$\mu$m. 
A notable feature is the rapid increase of the trap frequency along the $y$ axis as the fibers are brought close to the trap center.
This increase is because the positive charging make the potential steeper along the $y$ axis; negative surface charges would induce the opposite effect. 
The trapping potential for a 500~$\mu$m long cavity is found to have frequencies of $(4.4, 2.7, 3.8)$ MHz.
The most shallow trap depth is 0.13 eV along the $+z$ direction.


We would anticipate problems if the number of charges were slowly and randomly varying, as this would result in a time-dependent potential seen by the ion. 
We have previously observed timescales of such drifts to be from hours to days~\cite{Ong-NJP18}. 
This problem could be solved either by employing a discharging method~\cite{Pollack-PRD10} or by applying compensation voltages to the metal coating.
As long as the trap frequency can be monitored on time scales faster than those of the charge drifts, a feedback voltage can be applied to the metal coating to stabilize the trap frequency.


\begin{figure*}
\includegraphics[width=6.3in]{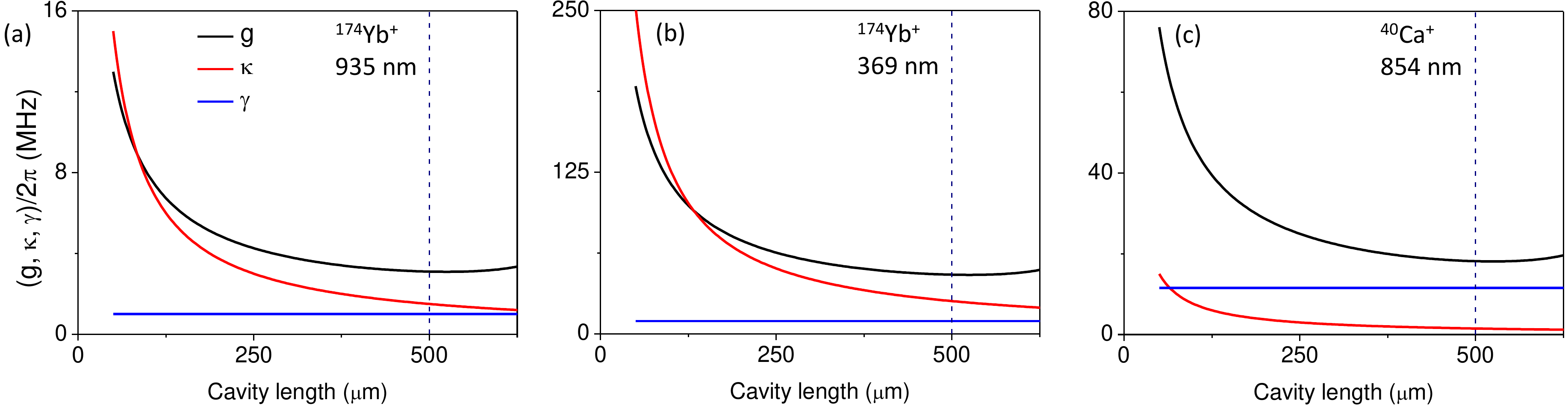} 
\caption{
(a) Calculated values for $g$ (black) and $\kappa$ (red) are plotted with with the atomic linewidth $\gamma$ (blue) for the $^{174}$Yb$^{+}$ $|^{2}D_{3/2}\rangle$--$|^{3}D[3/2]_{1/2}\rangle$ transition at 935 nm.
(b) The same parameters are shown for the $^{174}$Yb$^{+}$ $|^{2}S_{1/2}\rangle$--$|^{2}P_{1/2}\rangle$ transition at 369 nm, and (c) for the $^{40}$Ca$^{+}$ $|^{2}D_{5/2}\rangle$--$|^{2}P_{3/2}\rangle$ transition at 854 nm.
The color legends of (b) and (c) are identical to that of (a).
The dashed vertical lines indicate a cavity length of 500~$\mu$m, of particular relevance for our design (see text for details).
}
\label{fig:coop}
\end{figure*}

\section{Strong ion-cavity coupling}


Strong atom-cavity coupling means that the atom-cavity coupling constant $g$, i.e., half of the coherent energy exchange rate between the atom and the cavity field, is larger than both the atomic decay rate $\gamma$ and the cavity field decay rate $\kappa$~\cite{Kimble-PS98}.
The hallmark of strong atom-cavity coupling is the vacuum Rabi splitting, which originates from new atom-cavity eigenstates formed by the coupling~\cite{Haroche06}.
This effect offers capabilities for exploring fundamental quantum optics~\cite{Mabuchi2002} and implementing quantum network protocols~\cite{Reiserer-RMP15}.
Furthermore, if strong coupling can be achieved, then by choosing a different cavity geometry, one also has access to the Purcell regime, in which $\kappa \gg g^2/\kappa \gg \gamma$~\cite{Law1997,Reiserer-RMP15}.  The Purcell regime enables efficient single-photon sources, a key ingredient for constructing a modular ion-based quantum computer architecture~\cite{Monroe-PRA14, Kim-PRA11, Hucul-NatPHy15}.

In order to check whether strong ion-cavity coupling can be achieved with our design, we first plot calculated values of $g$ and $\kappa$ in Fig.~\ref{fig:coop}(a), with $\gamma$ given by the $^{174}$Yb$^{+}$ $|^{2}D_{3/2}\rangle$--$|^{3}D[3/2]_{1/2}\rangle$ transition at 935 nm.
The radii of curvature of both fiber mirrors are fixed at 350~$\mu$m and the effective diameter, i.e., the diameter that is used to form a cavity mode, at 100~$\mu$m~\cite{Ott-OE16}.
We assume a value of 97,000 for the cavity finesse, as reported in Ref.~\cite{Ott-OE16} for a recent fiber-cavity experiment. The finesse is a measure of the mirror losses and, together with the cavity length, allows $\kappa$ to be determined.
For a cavity length of 500~$\mu$m, the calculation results in $(g, \kappa, \gamma)/2\pi=(3.0, 1.0, 1.5)$ MHz, proving that the strong coupling condition can be satisfied for a length on which we have focused in the trapping potential calculations.

In Fig.~\ref{fig:coop}(b), we consider the $^{174}$Yb$^{+}$ $|^{2}S_{1/2}\rangle$--$|^{2}P_{1/2}\rangle$ transition at 369 nm.
The only different assumption here is a lower cavity finesse of 10,000, as reported in Ref.~\cite{Cetina-NJP13} and attributed to oxygen depletion in the mirror coating~\cite{Cetina-NJP13, Gangloff-OE15, Ballance-PRA17}.
For a cavity length of 500~$\mu$m, we find that the strong coupling condition still holds, with $(g, \kappa, \gamma)/2\pi=(46, 25, 9.9)$ MHz.  Note that while the 935~nm transition is more suitable for optical interconnects due to the lower absorption in optical fiber at this wavelength, the 369~nm transition enables significantly higher cavity coupling rates.
	
Finally, in Fig.~\ref{fig:coop}(c), we investigate the $^{40}$Ca$^{+}$ $|^{2}D_{5/2}\rangle$--$|^{2}P_{3/2}\rangle$ transition at 854 nm.
Assuming the same finesse, radii of curvature, and effective diameter as in the calculation shown in Fig.~\ref{fig:coop}(a), we obtain $g/2\pi=18$ MHz, which is greater than both $\kappa/2\pi=1.5$ MHz and $\gamma/2\pi=11.5$ MHz at a cavity length of 500 $\mu$m.
Taken together, the calculation results shown in Fig.~\ref{fig:coop} verify that our design can provide strong coupling for the transitions used in recent ion-fiber cavity experiments~\cite{Cetina-NJP13, Steiner-PRL13, Steiner-PRL14, Ballance-PRA17, Takahashi-PRA17, Takahashi-arXiv18}.
Although the simulations of our ion trap design described in the previous sections were carried out for $^{174}$Yb$^{+}$ ions, it would be straightforward to adapt them for $^{40}$Ca$^{+}$ ions.

\section{Discussion}

As presented thus far, our miniaturized ion-cavity setting exhibits safe fiber-mirror support with a large stroke, a stable ion trapping potential, and feasible strong ion-cavity coupling.
Moreover, our approach of combining a segmented ion trap with a cavity provides an additional advantage for optimizing alignment of the positions of multiple ions to the cavity field.
Suppose that multiple ions are coupled to a cavity for a geometry in which the trap axis and the cavity axis are parallel to one another, as for example in Ref.~\cite{Cetina-NJP13}. 
Such a system would be useful for an efficient quantum network node based on collective interactions~\cite{Lamata-PRL11}.
One technical challenge in realizing this ion-cavity system consists of the incommensurability between the ion--ion distances and the spatial period of the cavity standing wave. 
While the ion--ion distances are determined by the harmonic potential and the repulsive Coulomb interaction between the ions, the cavity-field period corresponds to one half of the resonant wavelength.  
Therefore, the strength of the coupling of each ion to the cavity field will be different, resulting in a reduced collective coupling rate as compared to the case in which each ion is coupled to an antinode of the cavity standing wave.

With our segmented trap design, through a particular choice of electrode voltages, it is possible to create a periodic trapping potential such that ions are equally spaced along the cavity axis, in order to couple each ion with an individual antinode of the cavity standing wave~\cite{Home-QIC06}.
The number of possible potential wells increases as one makes use of more segmented electrodes. 
Alternatively, a box-like quartic potential can be engineered by adjustment of the electrode voltages~\cite{Lin-EPL09}. 
Here the ion--ion distance is predominantly determined by the repulsive Coulomb interaction, which aligns the ions equidistantly.
Note that in a recent demonstration, up to five ions were positioned near cavity field antinodes based on visibility measurements as a function of trap frequency~\cite{Begley2016}. 
This method, however, would be difficult to extend for controlling the positions of arbitrary numbers of ions.


In order to scale the ion-cavity system down even further, the surrounding elements of the setup can be also miniaturized~\cite{Kim-QIC09}, drawing on the technologies of microfabrication and MEMS in order to precisely integrate optical elements, photodetectors, and parts of the control electronics.
There have been extensive experimental efforts to combine optical elements with surface traps at the micro- or nanometer scale~\cite{Merrill-NJP11, Herskind-OL11, Rynbach-AO17}.
For example, optical fibers~\cite{VanDevender-PRL10}, micromirrors, and a photodetector have been integrated to collect the ion fluorescence~\cite{Eltony-APL13, Clark-PRAppl14}, MEMS mirrors have been used to address individual ions~\cite{Knoernschild-OE09, Crain-APL14}, and more advanced optical elements like diffractive optics have also been integrated on a nanometer scale~\cite{Young-AO14, Mehta-NatNano16}.

\section{Conclusion}


We have proposed and investigated a novel design for MEMS-based manipulation of a fiber cavity integrated with a surface ion trap. 
The fiber stages are shown to be mechanically stable to support the fiber mirrors.
The stages move continuously in three dimensions with strokes up to tens of micrometers.
The mechanical resonance frequencies of the stage and actuator are found to be on the order of kHz. 
Furthermore, we calculate a stable trapping potential generated for a 500~$\mu$m long cavity, which should allow us to achieve strong ion-cavity coupling. 
This design for an on-chip ion-cavity system constitutes a promising approach for constructing a scalable quantum network. 

\section*{Acknowledgments}


This work has been financially supported by the Austrian Science Fund (FWF) through Projects F4019-N23, V252, and M1964; by the U.S. Army Research Laboratory's Center for Distributed Quantum Information via the project SciNet: Scalable Ion-Trap Quantum Network, Cooperative Agreement No. W911NF15-2-0060; by the Innsbruck-Seoul ION Trap Project; 
and by the Ministry of Science and ICT (MSIT), Korea, under the Information Technology Research Center (ITRC) support program (IITP-2019-2015-0-00385), supervised by the Institute for Information \& Communications Technology Promotion (IITP).

\section*{Open data}

Our data are available at \href{https://doi.org/10.5281/zenodo.3697922}{https://doi.org/10.5281/zenodo.3697922}.

\bibliographystyle{apsrev4-1}
\bibliography{MEMS_references}

\clearpage

\setcounter{equation}{0}
\setcounter{figure}{0}
\setcounter{table}{0}
\renewcommand{\theequation}{S\arabic{equation}}
\renewcommand{\thefigure}{S\arabic{figure}}
\renewcommand{\bibnumfmt}[1]{[S#1]}

\begin{center}
\textbf{\large Supplemental Material}
\end{center}

\section*{Mass of the fiber stage}

Information on the materials and densities used for the fibers and stages is given in Table~\ref{tab:density}, and information on the masses is given in Table~\ref{tab:mass}. 
We assume that the part of the fiber that the stage has to support has a length of 1 cm, where we would install an additional stage for relieving a mechanical strain to the main fiber stage or applying an additional feedback voltage to stabilize the cavity frequency. 
The mass of glue is neglected~\cite{Goettsche-IJAMT07}.

\section*{Details of the trapping potential calculation}

\subsection*{Simulation parameters}
 
\begin{figure}
\includegraphics[width=3in]{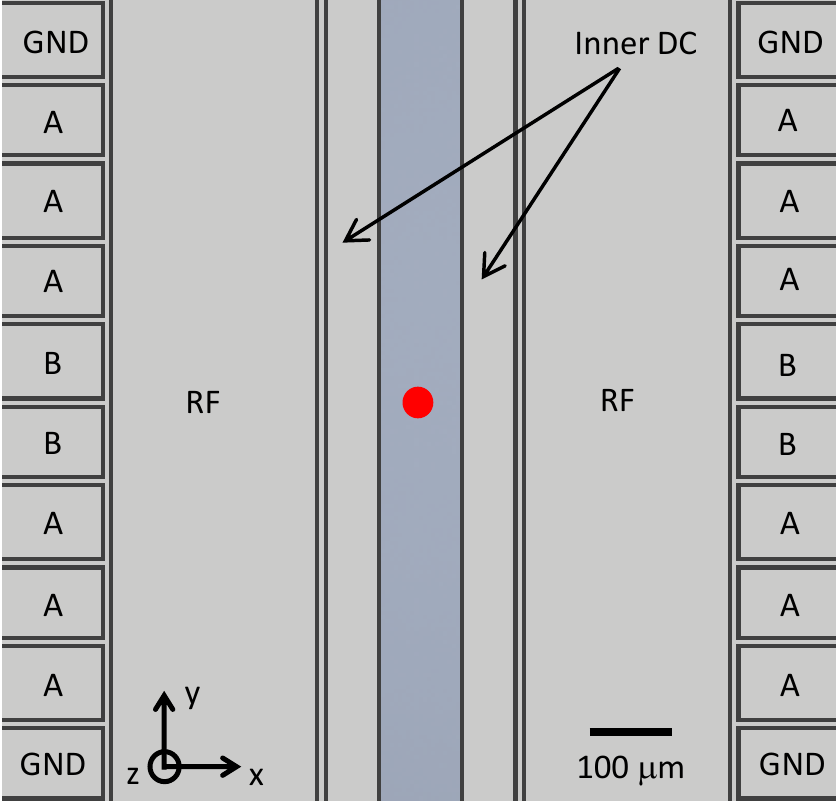} 
\caption{
The electrode configuration of the trapping chip.
The voltage $V_{\rm{A}}+V_{\rm{offset}}$ is applied to the outer DC electrodes A and the voltage $V_{\rm{B}}+V_{\rm{offset}}$ to electrodes B.
RF indicates the RF rail, and GND denotes ground. 
The red dot in the center indicates the ion position.
The voltage $V_{\rm{A}}+V_{\rm{offset}}$ is applied to the outer DC electrodes A and the voltage $V_{\rm{B}}+V_{\rm{offset}}$ to electrodes B.
RF indicates the RF rail, and GND denotes ground. 
The red dot in the center indicates the ion position.
}
\label{fig:electrodes}
\end{figure}

The parameters common to all simulations are an RF driving frequency of 20 MHz with an amplitude of 150 V, grounded inner DC electrodes, and a dielectric constant of the fiber (SiO$_2$) of 3.8. 
The parameters that are varied are given in Table~\ref{tab:voltage}.
Regarding the voltages applied to the outer DC electrodes (Fig.~\ref{fig:electrodes}), voltage $V_{\rm{A}}+V_{\rm{offset}}$ is applied to electrodes A and voltage $V_{\rm{B}}+V_{\rm{offset}}$ to electrodes B.
The difference $V_{\rm{A}}- V_{\rm{B}}$ determines the trap depth in the $y$ direction, and the offset voltage $V_{\rm{offset}}$ shifts the minimum of the DC potential in the $z$ direction and is chosen so that the minimum of the DC potential overlaps with that of the pseudopotential.

\subsection*{Micromotion compensation}

\begin{figure}
\includegraphics[width=3in]{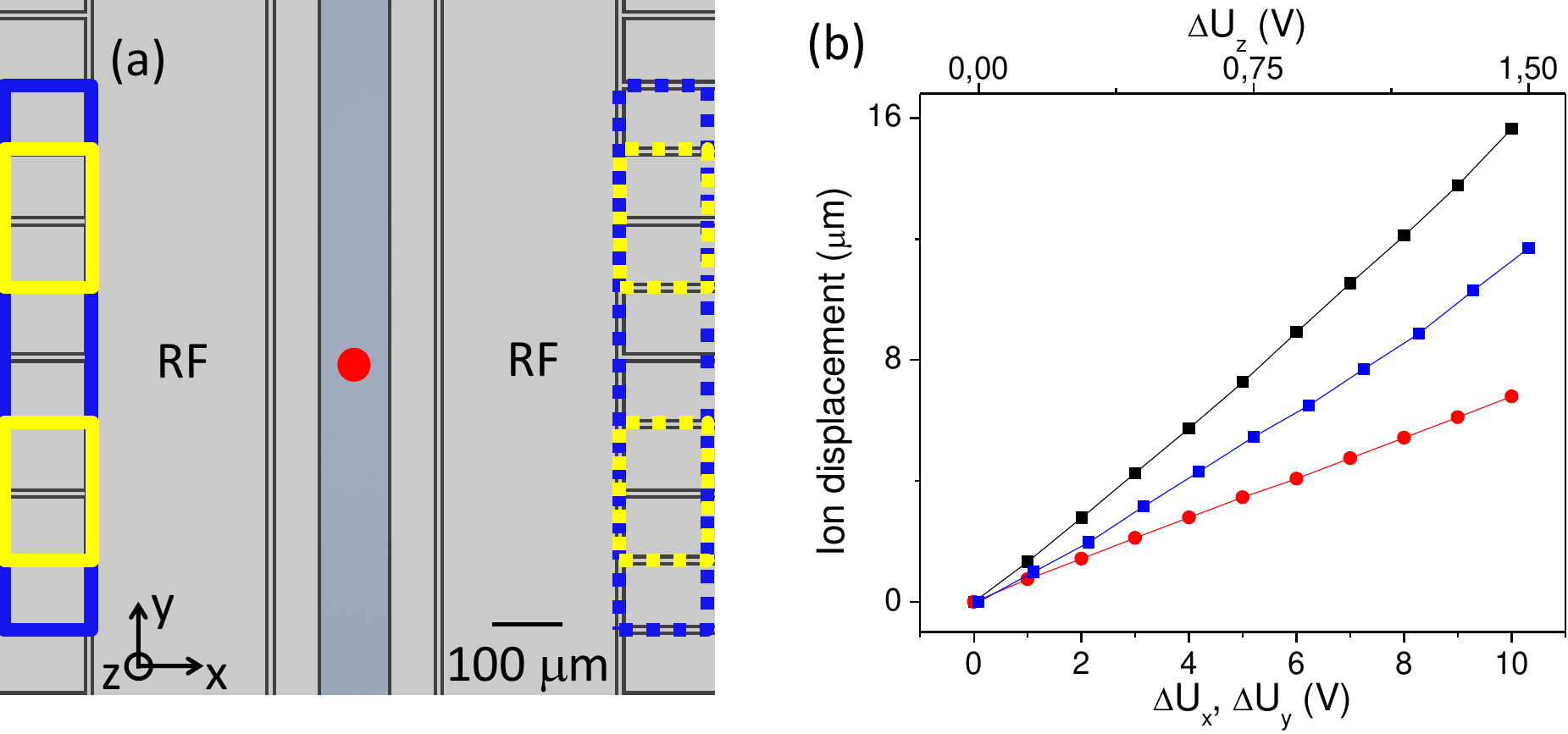} 
\caption{
Micromotion compensation.
(a) The outer DC electrodes in the blue boxes are used for compensation along the $x$ axis, and the electrodes in the yellow boxes are used for compensation along the $y$ axis.
The red dot in the center denotes the ion position.  
(b) Simulation results. 
The ion displacement along both axes is plotted as a function of the applied compensation voltages $\Delta U_x$ and $\Delta U_y$ (see text for definition). 
The $x$ displacement is shown in black and the $y$ displacement in red.
(a) The outer DC electrodes in the blue boxes are used for compensation along the $x$ axis, and the electrodes in the yellow boxes are used for compensation along the $y$ axis.
The red dot in the center denotes the ion position.  
(b) Simulation results. 
The ion displacement along both axes is plotted as a function of the applied compensation voltages $\Delta U_x, \Delta U_y,$ and $\Delta U_z$ (see text for definition). 
The $x$ displacement is shown in black, the $y$ displacement in red, and the $z$ displacement in blue.
}
\label{fig:micromotion}
\end{figure}

It is possible to compensate micromotion of the ions via tuning the voltages applied to the outer DC electrodes (Fig.~\ref{fig:micromotion}(a)).
For example, in order to shift the ion position in the $+x$ direction, the voltage applied to the DC electrodes marked with blue solid lines has to be changed by $\Delta U_x$ while the voltage applied to the electrodes marked with dashed blue lines is changed by $-\Delta U_x$. 
Similarly, shifting the ion in the $+y$ direction by an amount $\Delta y$ requires voltage changes of $+\Delta U_y$ and $- \Delta U_y$ on the electrodes marked by solid and dashed yellow lines, respectively. 
The ion position moves along the $+z$ direction by applying $+\Delta U_{z}$ to all outer DC electrodes.
The simulation result, displayed in Fig.~\ref{fig:micromotion}(b), shows that the ion position can be displaced by a few microns in each direction with voltage changes of only a few volts. 
This is expected to be sufficient to bring the ions to the RF-null point for typical DC background fields.

\section*{Fabrication strategy}

\begin{figure*} [!t]
\includegraphics[width=5.5in]{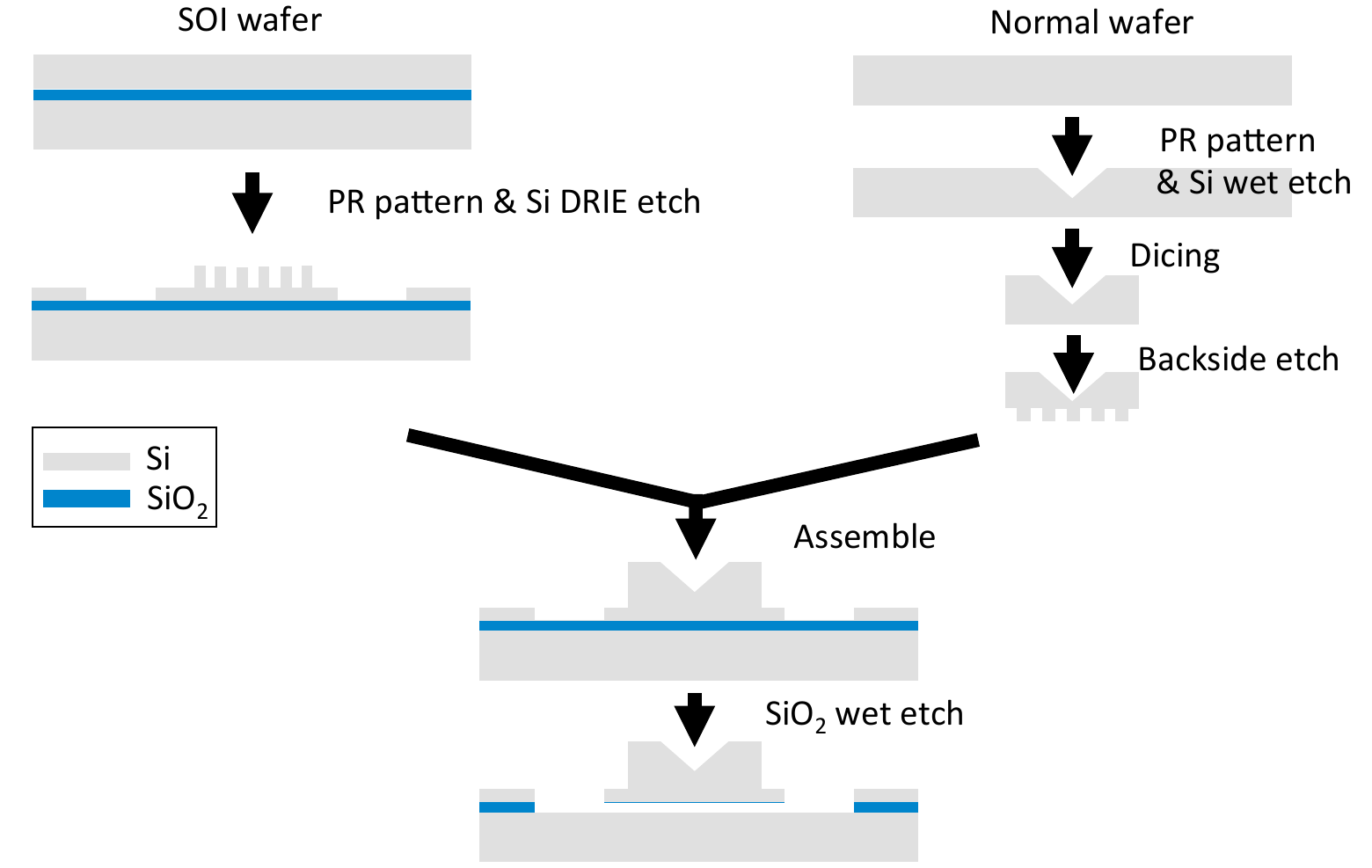} 
\caption{
Fabrication process of the fiber bench, stage, and actuator. 
}
\label{fig:fabrication}
\end{figure*}

The fabrication process of the MEMS-based fiber stage and actuator~\cite{Tang-SA89, Liu-JMM07, Maroufi2016} is shown in Fig.~\ref{fig:fabrication}.
The comb drive arrangement with its plug-and-socket structure is patterned on a silicon-on-insulator (SOI) wafer with a photoresist (PR) method~\cite{Diem-SAA95, Renard2000}.
The patterned structure is fabricated with the technique of silicon deep reactive ion etching (DRIE). 
In parallel with this process, the V-shaped fiber bench is prepared in the following way. 
The V-groove is shaped with PR patterning and silicon wet etching. 
The fiber bench is diced out afterwards with the correct dimension, followed by PR pattering and DRIE for the plug-and-socket structure on the backside. 
Next, these two parts, the comb drive and the fiber bench, are plugged and glued. 
Note that the plug-and-socket structure defines the position with a precision of few micrometers.
After this assembly, SiO$_{2}$ wet etching follows to complete the development of the fiber stage~\cite{Kim-Trans03}.
The fiber stage structure can also be fabricated by the surface/bulk micromachining process~\cite{Lee-JMS99}.


\begin{table*}  
\begin{tabular}{l*{4}{c}r} 
                    && Material && Density (g/cm$^{3}$) \\
\hline
Fiber              && SiO$_{2}$ && 2.65 \\
Fiber stage      && Si && 2.57  \\
Actuator          && Si && 2.57 \\
\end{tabular}

\caption{The materials and densities of the dielectrics used in the simulations of mechanical properties.}

\label{tab:density}
\end{table*}

\begin{table*} 
\begin{tabular}{l*{4}{c}r} 
                    && Mass ($\mu$g) \\
\hline
Fiber stage                && 235  \\
...with bench                && 806    \\
...with bench and fiber  && 1806 \\
\end{tabular}
\caption{Estimated mass of the fiber stage elements used in the simulations of mechanical properties.}
\label{tab:mass}
\end{table*}

\begin{table*}  
\begin{tabular}{l*{8}{c}r} 
                    &&  $V_{\textrm{A}}$ (V) && $V_{\textrm{B}}$ (V) && $V_{\textrm{offset}}$ (V) \\
\hline
Uncharged and uncoated         && 24.3 && -65.7 && 0.0\\
Uncharged and coated            && 26.6 && -63.5 && 0.0\\
Charged and coated                && 19.4  && -70.7 && \textrm{Varied from -24 to +3 V for each cavity length}\\
\end{tabular}
\caption{RF amplitudes and DC voltages used in the simulations of the trapping potential in Sec.~\rom{4} of the main text. ``(Un)charged" refers to the charge state of the dielectric fiber facets, and ``(un)coated" refers to presence or absence of a metal coating on the fiber circumference.
}
\label{tab:voltage}
\end{table*}

\end{document}